\documentclass{ws-procs9x6}

\begin{document}

\title{A FUNCTIONAL RENORMALIZATION GROUP APPROACH TO
SYSTEMS WITH LONG-RANGE CORRELATED DISORDER}

\author{ANDREI A. FEDORENKO}

\address{CNRS-Laboratoire de Physique Th{\'e}orique de l'Ecole Normale Sup{\'e}rieure,\\
 24 rue Lhomond, 75231 Paris,  France\\
E-mail: andrei.fedorenko@lpt.ens.fr}

\begin{abstract}
We studied the statics and dynamics of
elastic manifolds in  disordered media with long-range correlated disorder
using functional renormalization group (FRG).
We identified different universality classes
and computed the critical exponents and universal amplitudes
describing geometric and  velocity-force characteristics.
In contrast to uncorrelated disorder, the statistical
tilt symmetry is broken  resulting in a nontrivial response to a transverse tilting
force. For instance, the vortex lattice
in disordered superconductors  shows a new glass phase whose properties interpolate
between those of the Bragg and Bose glasses formed by pointlike and columnar
disorder, respectively. Whereas there is no response in  the Bose glass phase
(transverse Meissner effect), the standard linear response expected in the Bragg
glass  gets modified to a power law response in the presence of  disorder correlations.
We also studied the long distance properties of the $O(N)$ spin system  with
random fields and random anisotropies correlated as $1/x^{d-\sigma}$.
Using FRG we obtained the phase diagram in $(d,\,\sigma,\,N)$-parameter space and computed
the corresponding critical exponents. We found that below the lower critical
dimension $4+\sigma$, there can exist two different types of quasi-long-range-order with
zero order-parameter but infinite correlation length.
\end{abstract}


\bodymatter

\section{Elastic manifolds in disordered media}
Elastic objects in disordered media are a fruitful concept to study diverse physical systems
such as domain walls (DW) in ferromagnets, charge density waves (CDW) in solids, vortices
in type-II superconductors.\cite{aaf:ref0} In all these systems the
interplay between elastic forces which tend to keep the system
ordered, i.e., flat or periodic, and quenched disorder, which
promotes deformations of the local structure, forms a complicated
energy landscape with numerous metastable states.
Most studies of elastic objects in  disordered media
restricted to uncorrelated pointlike disorder. Real systems, however,
often contain extended defects in the form of linear dislocations, planar grain
boundaries, three-dimensional cavities, etc.
We studied the statics and dynamics of elastic
manifolds in the presence of (power-law) long-range (LR) correlated
disorder.\cite{aaf:ref1}
The power-law correlation of defects in  $d$-dimensional space with exponent
$a=d-\varepsilon_d$  can be ascribed to randomly distributed
extended defects of internal dimension $\varepsilon_d$ with random
orientation.

The configuration of elastic object can be parametrized by  displacement
field $u_{x}$, where $x$ denotes the $d$ dimensional internal coordinate of
the elastic object. The Hamiltonian is given by
\begin{equation}
  \mathcal{H}[u] = \int d^d x \left[ \frac{c}2 (\nabla u_x)^2 + V(x,u_x)  \right ].
\end{equation}
$c$ is the elasticity, $V(x,u)$ a random potential
with zero mean and variance
\begin{eqnarray}
  \overline{V(x,u)V(x',u')}&=&R_1(u-u')\delta^d(x-x')
+ R_2(u-u')g(x-x'),
\end{eqnarray}
where $g(x)\sim x^{-a}$.
For interfaces one has to  distinguish two universality classes:
random bond (RB) disorder described by a short-range function $R(u)$ and
random field (RF) disorder corresponding to a function which behaves as $R(u) \sim |u|$
at large $u$. Random periodic (RP) universality class corresponding to a periodic
function $R(u)$ describes systems such as CDW or vortices in $d=1+1$
dimensions.
Disorder makes the interfaces rough
with displacements growing with the distance $x$ as
$C(x)\sim x^{2\zeta}$, where $\zeta$ is the roughness exponent.
Elastic periodic structures lose their strict translational order
and may exhibit a slow logarithmic growth of
displacements,  $C(x) = \mathcal{A}_d \ln |x|$.
The driven dynamics at zero
temperature is described  by the over-damped equation of motion
\begin{equation}
\eta\partial_t u_{x t} = c\nabla^2 u_{x t} + F(x,u_{x t})+f.
\end{equation}
Here $\eta$ is the friction coefficient, $f$ the applied force
and $F=-\partial_u V(x,u) $ the pinning force with zero mean and correlator
\begin{eqnarray}
\overline{F(x,u)F(x',u')}&=&{\Delta}_1(u-u')\delta^d(x-x') + {\Delta}_2(u-u')g(x-x'),
\end{eqnarray}
where $\Delta_i(u)=-R_i^{\prime\prime}(u)$. The system undergoes
the so-called depinning transition at the critical force $f_c$,
which separates sliding and pinned states.
Upon approaching the depinning transition from the sliding state $f \to f_c^+$
the center-of-mass velocity $v=L^{-d}\int_x\partial_t u_{xt}$
vanishes as a power law $ v \sim (f-f_c)^{\beta}$ and the time and length
scales are related by $t \sim x^z$.

Let us  firstly discuss the case of short range (SR) correlated
disorder ($\Delta_2$=0). The problem turns out to be notably difficult
due to  the  so-called dimensional reduction (DR). However,
simple Imry-Ma arguments show that DR is wrong.
The metastability renders the zero temperature perturbation theory
useless breaking it down on scales larger than the so-called
Larkin length. To overcome these difficulties one can apply
functional renormalization group (FRG).  Scaling analysis gives that the large-scale
properties of a $d$-dimensional  elastic system are governed by uncorrelated disorder
in $d<d_{\mathrm{uc}}=4$.  The peculiarity of the problem is that
there is an infinite set of relevant operators which
can be parametrized by a function. The latter is nothing but the disorder correlator.
We generalized the FRG approach to systems with LR correlated disorder.
Using a double expansion in $\varepsilon=4-d$ and $\delta=4-a$ we derived the
flow equations to one-loop order:
\begin{eqnarray}
\partial_{\ell}  {\Delta}_1(u) &=&(\varepsilon-2\zeta)  {\Delta}_1(u)
+\zeta u  {\Delta}_1'(u) -
\frac12 \frac{d^2}{d u^2}[ {\Delta}_1(u)+ {\Delta}_2(u)]^2
 + A {\Delta}_1''(u), \nonumber  \\
\partial_{\ell}  {\Delta}_2(u) &=& (\delta-2\zeta) {\Delta}_2(u) +
 \zeta u  {\Delta}_2'(u)
  + A  {\Delta}_2''(u),\qquad
\label{aaf:frg-del-2}
\end{eqnarray}
where $A = [ {\Delta}_1(0) + {\Delta}_2(0)]$.
The scaling behavior of the system
is controlled by a stable fixed point $[{\Delta}_1^*(u), {\Delta}_2^*(u)]$ of
flow equations (\ref{aaf:frg-del-2}).
Note that for uncorrelated disorder the elasticity remains
uncorrected to all orders due to the statistical tilt symmetry (STS).
LR disorder destroys the STS and allows
for the renormalization of elasticity introducing  a new exponent $\psi$.
The SR part of disorder correlator $\Delta_1(u)$ becomes a
nonanalytic function beyond the Larkin scale, while the
LR part $\Delta_2(u)$ remains analytic along the flow.
The appearance of non-analyticity in the form of a cusp at the origin is related to
metastability, and nicely accounts for the generation of threshold
force $f_c$ at the depinning transition.
The actual values of critical exponents are fixed by the disorder correlator at
the stable FP, for instance,
$\psi = \frac14 (\delta-\varepsilon ) \Delta^{*\prime\prime}_2(0)$
and $z= 2- {\Delta}_1^{*\prime\prime}(0)-{\Delta}_2^{*\prime\prime}(0)$. Let us
summarize the results obtained for different universality classes.
\paragraph{RB disorder.} The crossover from the SR FP ($\Delta_2=0$) to the
LR FP ($\Delta_2\ne 0$) takes place for ${\delta}> 1.041\varepsilon$.
At the LR RB FP there is an exact relation between exponents:
$\zeta_{\mathrm{LR}}=(\delta+2\psi)/5$, where
$\psi=\mathcal{O}(\varepsilon^2,\delta^2,\varepsilon\delta)$ to one-loop order.
We also computed the amplitude of  height-height  correlation function which turns out to
be universal up to the strength of disorder.
\paragraph{RF disorder.} The large-scale properties of the system is controlled
by the LR FP for $\delta>\varepsilon$. The roughness exponent is given by the exact relation
$\zeta_{\mathrm{LR}}=(\delta+2\psi)/3$.
Surprisingly, the function $\Delta^*_1(u)$ satisfies
$\int_0^{\infty}du\ \Delta^*_1(u)=0$, characteristic for RB type correlations
along the $u$ direction.  In other words, the LR RF FP is in fact of
mixed type: RB for the SR part and RF for the LR part of the disorder
correlator. The RF FP describes the depinning transition of interfaces
in the presence of LR correlated disorder. The corresponding dynamic exponents read:
$z = 2-{\varepsilon}/3  + {\delta}/9$ and $\beta = 1-{\varepsilon}/6+ {\delta}/{18}$.
It is remarkable that for $\delta > 3\varepsilon$ the exponent
$\beta$ is larger than $1$, and $z$ larger than $2$.
\paragraph{Random periodic.} For $\delta>{\pi^2}{\varepsilon}/{9}$ the periodic
system is controlled by the LR RP FP. The system exhibits a slow growth of the
displacements:   $\overline{(u_x-u_0)^2}= \frac{\delta}{2\pi^2} \ln |x|.$
As an example, we consider vortices in the presence of three types of disorder:
uncorrelated, LR and columnar. The correlation of disorder violates the STS
resulting in a highly nonlinear response to tilt.
In the presence of columnar disorder vortices exhibit a transverse
Meissner effect: disorder generates the critical field $h_c$ below which
there is no response to tilt and above which the tilt angle behaves
as $\vartheta\sim(h-h_c)^{\phi}$ with the universal exponent $\phi<1$.
The RP case describes a weak Bose glass which is expected in type-II superconductors
with columnar disorder at small temperatures and at high vortex density which exceeds
the density of columnar pins. The weak Bose glass is pinned collectively and
shares features of the Bragg glass, such as a power-law decay of the translation order,
and features of the strong Bose glass such as a transverse Meissner effect.
For isotropically LR correlated disorder the linear tilt modulus vanishes
at small fields leading to a power-law response $\vartheta\sim h^{\phi}$
with $\phi>1$. The response of the system with LR correlated disorder
interpolates between responses of systems with uncorrelated and columnar disorder.
We argued that in the presence of LR correlated disorder vortices
can form a strong Bragg glass which exhibits Bragg peaks and vanishing
linear tilt modulus without transverse Meissner effect.\cite{aaf:ref2}

\section{LR random field and random anisotropy $O(N)$ models}

\begin{figure}[tbp]
\includegraphics[clip,width=4.5in]{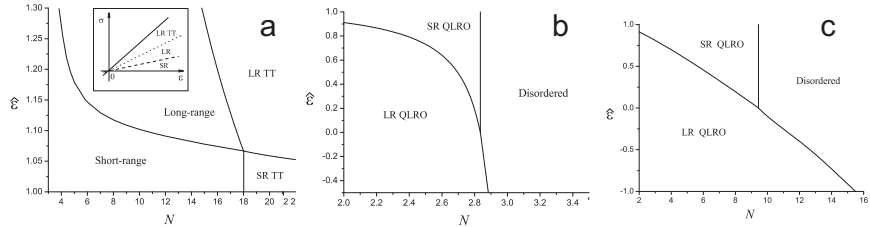}
\caption{ (a)~Stability regions of various FPs for RF $O(N)$ model
corresponding to different patterns of the critical behavior above
the lower critical dimension, ${\varepsilon}>\sigma$. Insert: Schematic phase
diagram on the $(\varepsilon,\sigma)$ plane for a particular value of $N\in(3,18)$.
(b)~Phase diagram of the RF model below the lower critical
dimension, $d<4+\sigma$.
(c)~Phase diagram of the RA model
below the lower critical dimension, $d<4+\sigma$.  }
\label{aaf:fig1}
\end{figure}

The large-scale behavior of the $O(N)$ symmetric spin system at low temperatures
can be described by the nonlinear $\sigma$ model with the Hamiltonian
\begin{eqnarray}
 \mathcal{H}\left[ \vec{s}\, \right] = \int d^d x \left[
   \frac{1}{2} (\nabla \vec{s}\,)^2 + V(x,\vec{s}) \right], \label{aaf:ham}
\end{eqnarray}
where $\vec{s}( x )$ is  the $N$-component classical spin  with a fixed-length constraint
$\vec{s}\,^2=1$.  $\mathrm{V}(x,\vec{s})$ is the random disorder potential, which can be
expanded in spin variables:
$V(x,\vec{s})= -\sum_{\mu=1}^{\infty}\sum_{i_1...i_ {\mu}}
h^{(\mu)}_{i_1...i_ {\mu}}(x) s_{i_1}(x)...s_{i_{\mu}}(x)$.
The corresponding coefficients have simple physical interpretation:
$h^{(1)}_i$ is a random field,
$h^{(2)}_{ij}$ is a random second-rank anisotropy, and $h^{(\mu)}$ are
general $\mu$th tensor anisotropies.
We studied system (\ref{aaf:ham})
with LR correlated disorder given by cumulants
$\overline{h^{(\mu)}_{i_1...i_{\mu}}(x) h^{(\nu)}_{i_1...j_{\nu}}(x')}=
 \delta^{\mu\nu}\delta_{i_1 j_1}
...\delta_{i_{\mu} j_{\nu}} [r_1^{(\mu)}\delta(x-x') + r_2^{(\mu)}g(x-x')]$
with $g( x ) \sim |x|^{\sigma-d}$.
The corresponding replicated Hamiltonian reads\cite{aaf:ref3}
\begin{eqnarray}
\mathcal{H}_n &=&
   \int d^d x \left[
   \frac{1}{2} \sum_{a} (\nabla \vec{s}_a \,)^2
 - \frac{1}{2T} \sum_{a,b} R_1\big(\vec{s}_{a}( x )\cdot\vec{s}_{b}( x )\big) \right. \nonumber \\
&& - \left. \frac{1}{2T} \sum_{a,b}  \int d^d x'\, g( x - x' )\,
 R_2\big(\vec{s}_{a}( x )\cdot\vec{s}_{b}( x' )\big)
    \right],\ \ \ \ \
\end{eqnarray}
where $R_i(z)=\sum_{\mu}r_i^{(\mu)} z^{\mu}$
are arbitrary for RF and even for RA.
Power counting suggests that $d_{\mathrm{lc}}=4+\sigma$ is the lower
critical dimension for both models.
At criticality  or in the quasi-long-range-ordered  (QLRO) phase the
connected and disconnected two-point function scale with different exponents
$\eta$ and $\bar{\eta}$.
The FRG equations to first order in
$\varepsilon=4-d$ and $\sigma$ are given  by
\begin{subequations}
\begin{eqnarray}
    \partial_{\ell} R_1(\phi)
        &=& - \varepsilon R_1(\phi) + \frac12 \big[R_1''(\phi) + R_2''(\phi) \big]^2
           - (N-2) \bigg\{ 2A R_1(\phi)  \nonumber \\
        &&   -A R_1'' (\phi) + A R_1'(\phi)\cot\phi
          - \frac1{2\sin^2{\phi}} \big[ {R_1'(\phi) + R_2'(\phi) } \big]^2 \bigg\}, \nonumber  \\
    \partial_{\ell} R_2(\phi)
        &=& - (\varepsilon-\sigma) R_2(\phi) - \Big\{ (N-2)\big[2R_2(\phi)
         + R_2'(\phi) \cot\phi\, \big] +  R_2''(\phi)\Big\} A. \nonumber
\end{eqnarray}
\end{subequations}
Here $A = R_1''(0)+R_2''(0)$ and $z=\cos\phi$.
An attractive FP of the flow equations describes a QLRO phase,
while a singly unstable FP describes the critical behavior.
The critical exponents are determined by the FRG flow in the vicinity of the FP and
to one-loop order are given by $\eta =  - A^*$ and $\bar{\eta}=-\varepsilon - (N-1)A^*$.
A singly unstable FP has only one positive eigenvalue $\lambda_1$
 which determines the third independent exponent $\nu = 1/\lambda_1$
characterizing the divergence of the correlation length
at the transition.
For the RF model the critical exponents are given by
\begin{eqnarray} \label{rf-exp}
  \eta_{\mathrm{LR}} = \frac{\varepsilon-\sigma}{N-3}, \  \  \  \  \
  \bar{\eta}_{\mathrm{LR}} = \frac{2\varepsilon-(N-1)\sigma}{N-3}.
\end{eqnarray}
The stability regions of different FPs and the phase diagrams for the RF and
RA models are shown in Figure~\ref{aaf:fig1}.
For the LR RF model we found that the
truncated RG developed in work \cite{aaf:chang84} for the RF model
gives the correct one-loop values of exponents $\eta$ and
$\bar{\eta}$ above the $d_{\mathrm{lc}}$, but not the phase diagram
and the critical exponent $\nu$ except for the region controlled by
the weakly nonanalytic LR TT FP.  Thus, although the truncated RG overcomes the
dimensional reduction it fails to reproduce all properties which can be
obtained using FRG. We found a new LR QLRO phase
existing in the LR RF model below the $d_{\mathrm{lc}}$
for $N<3$ and determined the regions of its stability in the
$(\varepsilon,\sigma,N)$ parameter space. We obtained
that the weak LR correlated disorder does not change the critical behavior
of the RA model above $d_{\mathrm{lc}}$ for $N>N_c=9.4412$, but can create
a new LR QLRO phase below $d_{\mathrm{lc}}$. The existence of two QLRO phases
in the LR RA system may be relevant to two different states of $^3$He-A in aerogel
observed recently in NMR experiments.\cite{aaf:volovik06}

\section*{Acknowledgments}
I would like to thank
Kay Wiese, Pierre Le Doussal and Florian K\"uhnel
for fruitful collaboration and useful discussions.
Support from the European Commission through a Marie
Curie Postdoctoral Fellowship under contract number
MIF1-CT-2005-021897 is acknowledged.

\end{document}